\documentclass[prb,twocolumn,showpacs,preprintnumbers,amsmath,amssymb]{revtex4}

\usepackage{graphicx}% Include figure files
\usepackage{amsmath}
\usepackage{epstopdf}
\usepackage{amsbsy}
\usepackage{dcolumn}% Align table columns on decimal point
\usepackage{bm}% bold math

\newcommand{\BSCCO}{{Bi$_2$Sr$_2$CaCu$_2$O$_{8+x}$ }}
\newcommand{\BSCCOoneplane}{{Bi$_2$Sr$_2$CuO$_6$ }}
\newcommand{\bea}{\begin{eqnarray}}
\newcommand{\eea}{\end{eqnarray}}
\newcommand{\be}{\begin{equation}}
\newcommand{\ee}{\end{equation}}
\def\k{{\bf k}}
\def\rr{{\bf r}}

\begin{document}

\title{Extinction of impurity resonances in large-gap regions of inhomogeneous $d$-wave superconductors}

\author{Brian M. Andersen$^1$, S. Graser$^2$, and P. J. Hirschfeld$^2$}

\affiliation{$^1$Nano-Science Center, Niels Bohr Institute,
University of Copenhagen, Universitetsparken 5, DK-2100 Copenhagen,
Denmark\\
$^2$Department of Physics, University of Florida, Gainesville,
Florida 32611-8440, USA}

\date{\today}

\begin{abstract}

Impurity resonances observed by scanning tunneling spectroscopy
in the superconducting state have been used to deduce properties
of the underlying pure state.  Here we study a longstanding puzzle
associated with these measurements, the apparent extinction of
these resonances for Ni and Zn impurities in large-gap regions of
the inhomogeneous \BSCCO superconductor.  We calculate the effect
of order parameter and hopping suppression near the impurity site,
and find that these two effects are sufficient to explain the
missing resonances in the case of Ni.  There are several
possible scenarios for the extinction of the Zn resonances, which
we discuss in turn; in addition, we propose measurements which
could distinguish among them.

\end{abstract}

 \maketitle
\section{Introduction}

An understanding of the low doped Mott region of cuprate
superconductors remains a key problem in condensed matter
physics.\cite{leereview} It has been proposed that electronic
inhomogeneity is  crucial for explaining experiments performed
in the underdoped regime,\cite{sakivelson} and that the
inhomogeneity arises from strong Coulomb repulsion which
instigates the formation of spin- and charge density waves, enhanced
and pinned by a random potential generated  by the doping process, or
by external probes which break the translational symmetry. Recent
theoretical modelling has shown that several single-particle and
two-particle experimental probes can be explained within a
scenario where the underdoped superconducting region consists of a
$d$-wave pair state coexisting with a local cluster spin glass
phase driven by disorder.
\cite{alvarez05,andersen07,atkinson07,andersenkappa,alvarez08}

Scanning tunneling microscopy/spectroscopy (STM/STS) measurements have provided a
wealth of detailed local density of states (LDOS) spectra on the
surface of \BSCCO (BSCCO), revealing an inhomogeneity in the
spectral gap on the nanometer scale.\cite{cren,howald,pan,lang}
Furthermore, the differential conductance has been measured in the
vicinity of Zn and Ni substituents for planar Cu.\cite{panzn,lang,hudson} Theoretically, the STS spatial
conductance pattern produced by the magnetic Ni ion and its
resonance energy of $\Omega_0\simeq \pm 10$ meV can be well
described by a combination of potential and magnetic
scattering,\cite{tangflatte} whereas no consensus has been reached
about the proper explanation of the pattern around the Zn
impurity, even though  the local perturbation caused by the ion
should in principle be
simpler.\cite{balatsky,vojta,alloul,andersen05}

One way to gain further insight into the nature of the physical
state of small-gap versus large-gap regions in cuprate
superconductors, is to compare the LDOS near Ni and Zn in these
two different environments. Experimentally, it is found that the
Ni resonances,  which are clearly distinguishable in the small-gap
regions, are completely absent in large-gap regions with $\Delta
\gtrsim 50 \mbox{meV}$ for optimally doped BSCCO.\cite{lang}  The
strength of the resonances which {\it are} observed does not
depend strongly on local gap size, but despite considerable noise
appears to anticorrelate slightly with the
gap.\cite{hoffmanthesis} It was speculated\cite{lang} that the
absence of Ni resonances in large-gap regions is due to the
distinct nature of the (pseudogap) phase characterizing these
domains. However, phase coherence is not necessary for the
creation of impurity resonances; the existence of the pseudogap in
the density of states itself should result in well defined
impurity resonant states, as found in Ref. \onlinecite{kruis}.
Indeed, recent STS measurements on native defects in underdoped
single-layer \BSCCOoneplane  samples have shown that the
resonances exist well above the superconducting critical
temperature $T_c$ into the pseudogap state.\cite{chatterjee} 

The LDOS near Zn is characterized by a sharp low-energy resonance
around $-2 \mbox{meV}$ in the small-gap regions of optimally doped
BSCCO.\cite{panzn} The resonance weight
and width depend strongly on the local environment,\cite{davis,hoffmanthesis} and, as in the Ni
case,\cite{lang} the resonances are never observed in the
large-gap regions.\cite{davis,hoffmanthesis} The temperature dependence of the Zn 
resonance has been measured within the superconducting state. 
It was found that the evolution of the peak in the range $30\mbox{mK}<T<52\mbox{K}<T_c$ is consistent with thermal broadening of the peak.\cite{kambara}
Within the nonlocal Kondo model for Zn impurities,\cite{vojta} it was recently argued that a
gap-dependent exchange coupling can lead to suppressed Zn
resonances in large-gap regions.\cite{kircan} However, this model
does not account for the actual spatial variation of the superconducting gap, and it is unclear how to explain the extinction of the Ni resonances.
In our view, the question of what causes the extinction of the
resonances and whether these effects can indeed be used as probes
of pseudogap physics\cite{kruis,hudson} is therefore still open.

For BSCCO materials, a significant part of the observed gap
inhomogeneity has been argued to originate from interstitial
oxygen dopant atoms;\cite{mcelroyscience} detailed theoretical
modelling of the LDOS spectra\cite{nunner1,nunner3} has led to the
remarkable conclusion that the pairing interaction itself seems to be
spatially varying in these materials on an atomic scale.  This
 scenario in which the pairing interaction is modulated locally by
 nearby dopant atoms also explained the origin of the
dominant so-called ${\mathbf{q}}_1$ peak in the Fourier
transformed scanning tunneling spectroscopy (FTSTS)
data,\cite{nunner2,mcelroyFTSTS} and the recently observed pair
density wave driven by the structural
supermodulation.\cite{slezak,andersenslezak} In addition, it
predicted the existence of  islands with finite gap above T$_c$,
as recently observed by Gomes {\it et al.}\cite{gomes} Such
islands lead naturally to broadened thermodynamic transitions
although this point remains controversial at
present.\cite{andersen06,loram}

Here, we show that within the modulated pairing scenario,  the
absence of Ni resonances in large-gap regions arises naturally
when one properly includes the nano-scale inhomogeneity of the superconducting
order parameter.  This occurs because in-gap impurity resonances
in a $d$-wave superconductor generically borrow their weight from
the coherence peaks.  In the large-gap regions, these peaks are
broad and weak, so they overlap and swamp the weaker resonances
located away from the Fermi level. In the small-gap regions, on
the other hand, the structures normally referred to as coherence
peaks are actually Andreev resonant states of the suppressed order
parameter,\cite{nunner1,acfang} and their height/width is significantly
larger/smaller than in the pure homogeneous case.  Thus in the small-gap regions the impurity
resonances are well-defined due to their smaller overlap with the continuum.
Within the modulated pairing scenario, which reproduces the
correlation of the coherence peaks' weight and position with the
dopant atoms imaged by STS,\cite{mcelroyscience} the presence or
absence of Ni resonances is therefore dependent on whether or not
interstitial out-of-plane oxygen dopants exist nearby.

\begin{figure}[b]
\includegraphics[clip=true,width=1.0\columnwidth]{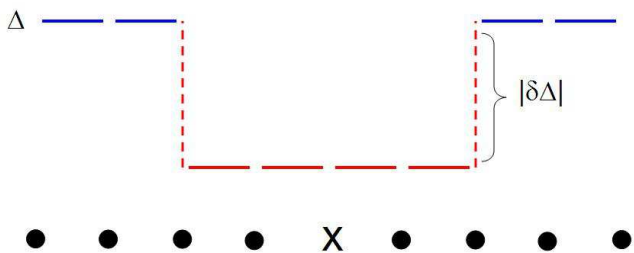}
 \caption{(Color online) Schematic picture of a small-gap region.  Circles are ionic sites, X represents an
 impurity, and the bonds represent the absolute values of the bond order parameter $\Delta$ on
 bonds in the bulk (top) and in a small gap patch (bottom). } \label{schematic}
\end{figure}

We further consider the effect of gap modulation on the Zn
resonance near the Fermi level.  Here it appears unlikely that the
suppression and broadening of the coherence peaks in large-gap
regions can alone suppress the impurity resonance because of its weak coupling to the continuum.  There is very little
experimental data on the variation of the spectral form, weight,
and position of the resonance in correlation with the local gap
size, but it appears from what is
available\cite{davis,hoffmanthesis} that there is considerable
dispersion of all these quantities which makes it difficult to
extract systematics.  Some of this dispersion may simply be due to
the well-known very strong interference of near-unitary impurity
resonances in a $d$-wave superconductor.\cite{stavropoulos,zah,andersen03} We
therefore consider a series of models for the microscopic nature
of the Zn potential which might possibly account for the
experimental observations, and propose additional experimental tests to
distinguish between them.

\section{Formalism}

The starting point of our study is given by the $d$-wave BCS Hamiltonian
\begin{equation}
{\mathcal{H}}_0=\sum_{{\mathbf{k}},\sigma} \xi_{\mathbf{k}}
\hat{c}^\dagger_{{\mathbf{k}}\sigma} \hat{c}_{{\mathbf{k}}\sigma}
+ \sum_{{\mathbf{k}}} \left( \Delta_{\mathbf{k}}
\hat{c}^\dagger_{{\mathbf{k}}\uparrow}
\hat{c}^\dagger_{-{\mathbf{k}}\downarrow} + {\rm H.c.} \right),
\end{equation}
where $\xi_{\mathbf{k}}$ denotes the quasiparticle dispersion and
$\Delta_{\mathbf{k}}=\frac{\Delta_0}{2}  (\cos k_x - \cos k_y)$ is
the $d$-wave pairing gap. In terms of the Nambu spinor
$\hat{\psi}^\dagger_{{\mathbf{k}}}=(\hat{c}^\dagger_{{\mathbf{k}}\uparrow},\hat{c}_{-{\mathbf{k}}\downarrow})$,
the corresponding Matsubara Green's function can be expressed as
\begin{equation}\label{nambugreensfunctions}
{\mathcal{G}}^{0}({\mathbf{k}},i\omega_n)=\frac{i\omega_n\tau_0+\xi_{\mathbf{k}}\tau_3
+\Delta_{\mathbf{k}}\tau_1}{(i\omega_n)^2-E_{\mathbf{k}}^2},
\end{equation}
where $E_{\mathbf{k}}^2=\xi_{\mathbf{k}}^2+\Delta_{\mathbf{k}}^2$,
and $\tau_i$ denote the Pauli matrices. In real-space, the
perturbation due to $\delta$-function potential (magnetic)
impurities of strength $V_3$ ($V_0$) in the diagonal $\tau_3$
($\tau_0$) channel is given by
\begin{equation}
{\mathcal{H}}^{\prime}_{imp}({\mathbf{r}},{\mathbf{r}}')=\hat{\psi}^\dagger_{{\mathbf{r}}}
\left[ \left( V_3\tau_3 + V_0\tau_0 \right)
\delta({\mathbf{r}})\delta({\mathbf{r}}') \right]
\hat{\psi}_{{\mathbf{r}}'}\label{potH}.
\end{equation}
Likewise, local modulations in the hopping ($\delta t$) or
superconducting gap ($\delta \Delta$) enter as
\begin{equation}
{\mathcal{H}}^{\prime}_{\delta}({\mathbf{r}},{\mathbf{r}}')=\hat{\psi}^\dagger_{{\mathbf{r}}}
\left[  -\delta t({\mathbf{r}},{\mathbf{r}}') \tau_3 - \delta
\Delta({\mathbf{r}},{\mathbf{r}}') \tau_1 \right]
\hat{\psi}_{{\mathbf{r}}'}.
\end{equation}

\begin{figure}[t]
\includegraphics[clip=true,width=1.0\columnwidth]{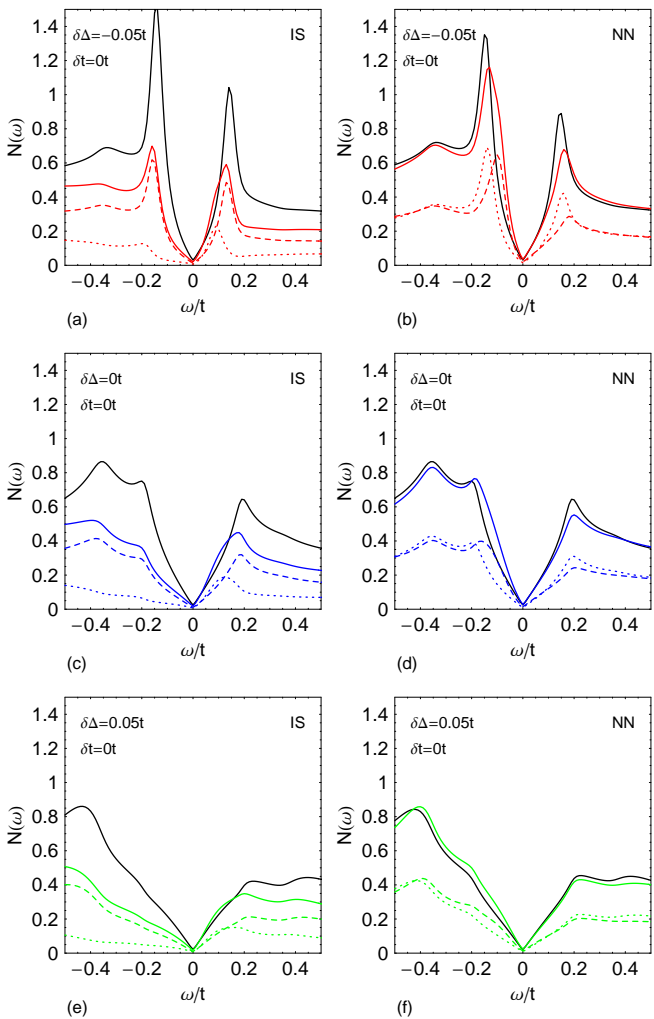}
\caption{(Color online) LDOS  in a $d$-wave superconductor
with a Ni impurity at the center of a 5$\times 5$ patch of
(a-b) suppressed ($\delta \Delta=-0.05t$), (c-d) constant ($\delta \Delta=0$) and (e-f)
enhanced ($\delta \Delta=+0.05t$)  superconducting gap. The Ni
potentials used for this study are $V_3=-0.8t$, $V_0=0.6t$.  The figures in the left
column (a,c,e) show spectra on the impurity site (IS) itself,
while right column (b,d,f) represents nearest neighbor (NN) sites.
In each case, the solid black curve shows the LDOS on the site in
question in the absence of the impurity, the solid color curve
indicates the total LDOS with the impurity, and the dashed and
dotted lines show the spin resolved LDOS for up and down spins,
respectively.} \label{fig1}
\end{figure}

To obtain the resulting LDOS as a function of position and energy, one needs to determine the full Green's function
${\mathcal{G}}({\mathbf{r}},i\omega_n)$ given by the Dyson
equation
\begin{equation}\label{dyson}
{\mathcal{G}}({\mathbf{r}},{\mathbf{r}}')={\mathcal{G}}^{0}({\mathbf{r}}-{\mathbf{r}}')+
{\mathcal{G}}({\mathbf{r}},{\mathbf{r}}'') {\cal
H}^{\prime}({\mathbf{r}}'',{\mathbf{r}}''')
{\mathcal{G}}^{0}({\mathbf{r}}'''-{\mathbf{r}}'),
\end{equation}
where ${\cal H}^\prime ={\cal H}^\prime_{imp}+{\cal
H}^\prime_\delta$, and  summation over repeated indices is
implied. Thus, by calculating the matrix elements of
${\mathcal{G}}^{0}({\mathbf{r}},i\omega_n)=
\sum_{\mathbf{k}}\mathcal{G}^{0}({\mathbf{k}},i\omega_n)e^{i
{\mathbf{k}} \cdot {\mathbf{r}}}$
the remaining problem is that of a matrix
inversion. The solution is presented in terms of the {$\cal T$}-matrix
\begin{equation}\label{tmat}
{\mathcal{G}}({\mathbf{r}},{\mathbf{r}}')={\mathcal{G}}^{0}({\mathbf{r}}-{\mathbf{r}}')+
{\mathcal{G}}^{0}({\mathbf{r}}-{\mathbf{r}}'') {\cal
T}({\mathbf{r}}'',{\mathbf{r}}''')
{\mathcal{G}}^{0}({\mathbf{r}}'''-{\mathbf{r}}').
\end{equation}
The poles of the ${\cal T}$-matrix, or equivalently, the zeros in
the determinant of $(1-{\cal H}^{\prime}{\cal G}^{0})$, determine
the bound state energies. The total (spin summed) LDOS is given by
\begin{equation}
N({\mathbf{r}},\omega)=(-1/\pi) \mbox{Im} \left[
{\mathcal{G}}_{11}({\mathbf{r}},{\mathbf{r}},\omega) +
{\mathcal{G}}_{22}({\mathbf{r}},{\mathbf{r}},-\omega)\right].
\end{equation}

\begin{figure}[t]
\includegraphics[clip=true,width=1.0\columnwidth]{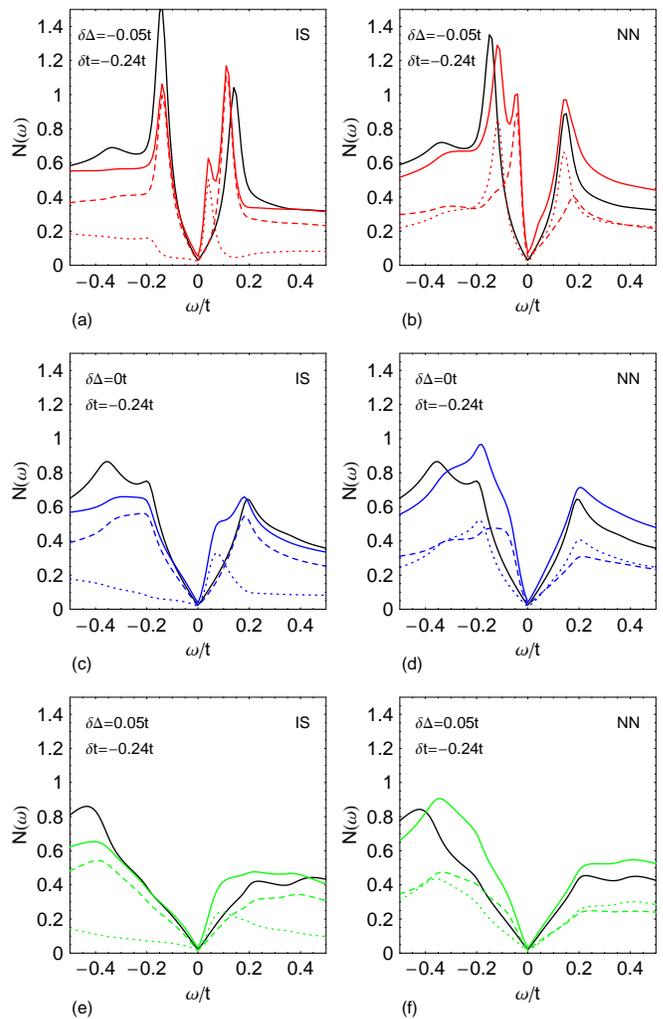}
 \caption{(Color online) Same as Fig. \ref{fig1} but including local impurity-induced reduced hopping $\delta t=-0.24t$.} \label{fig2}
\end{figure}

In explicit calculations we use the band $\xi_\k = -2t\,
(\cos k_x+\cos k_y) - 4t' \cos k_x \cos k_y - \mu$ with a $d$-wave
order parameter $ \Delta_\k = \frac{\Delta_0}{2}\ (\cos k_x-\cos
k_y)\, . $ To model the Fermi surface in the cuprates we take $t'
= -0.35 t$ and $\mu=-1.1 t$, with energy measured in units of
nearest neighbor hopping $t$. We also choose $\Delta_0=0.2t$.
The modulations in the hopping are included on the bonds near  the
impurity site assuming maximum modulations $\delta t$
on the nearest neighbor bonds, with a Gaussian decay on
other bonds with decay length of one lattice constant.  This is
roughly consistent with the renormalization of the effective
hopping found in recent studies of impurities in a host described by
the $t-t'-J$ model.\cite{gabay} Note this is an atomic scale modulation
caused by the in-plane impurity due solely to magnetic
correlations in the host. By contrast, we also study a modulation
of the order parameter, but assume that this is caused by an
out-of-plane influence such as a dopant atom, which gives rise to
a patch of modulated $\Delta_\k$ of roughly 20 \AA. Thus we
study the impurity added to a patch of reduced or enhanced
$\Delta_\k$ on a 5$\times$5 square as shown in Fig. \ref{schematic} embedded in an infinite system
of fixed order parameter $\Delta_0$.

\section{N\lowercase{i} impurity}

We first discuss the case of magnetic impurity ions such as Ni. In
Fig. \ref{fig1}  we show the LDOS at the magnetic impurity site
using the parameters $V_3=-0.8t$,$V_0=0.6t$,  and $\delta
\Delta=-0.05t, 0, +0.05t$ corresponding to a local suppressed,
constant, and enhanced gap patch, respectively.
The ${\cal T}$-matrix corresponding to this potential
has a pole at an energy $\Omega_0\sim0.05t$, but, as is evident from Fig. \ref{fig1},
 resonance are visible neither on the impurity
site (IS) nor on the nearest neighbor (NN) site.  Note that while
in Fig. \ref{fig1} we have assumed a quasiparticle scattering rate
$\Gamma(\omega)=0.1|\omega|$ similar to that observed in
experiment\cite{alldredge}, the absence of resonance features is
not caused by this broadening.  Instead, it is the fact that the
coupling to the continuum at $\Omega_0$ is simply too strong.  In
Fig. \ref{fig2}, we show the same sites with the same potentials
in a situation where the hopping has been reduced around the Ni
impurity; it is clear that the resonance is observable, and
corresponds  very closely in energy and weight to
experiment.\cite{hudson} It appears, however, only in the case
where the order parameter has been suppressed [Fig.
\ref{fig2}(a,b)] over a patch, thus creating the sharp coherence
peaks which are then separated from the impurity feature.  At this
point, the role of the suppressed hopping is purely
phenomenological, but we note that such a reduction was also
assumed in the more complicated  set of potential parameters taken
by Tang and Flatt\'e\cite{tangflatte} to describe the Ni
resonance. The particular values of $\delta t$ and $\delta \Delta$
necessary for the LDOS near the Ni impurity to resemble the
experimental data are bandstructure-dependent.

\begin{figure}[t]
\begin{center}
 \includegraphics[clip=true,width=.9\columnwidth]{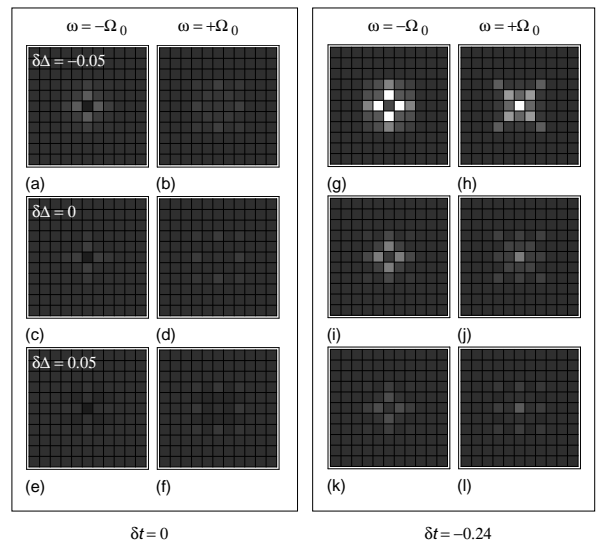}
\caption{LDOS real-space pattern $N({\bf r},\pm \Omega_0)$ at the
resonance frequency for Ni impurities modelled as described in the text.  Left panel: no
suppression of hopping by impurity assumed.  Subpanels (a-b), (c-d),
and (e-f) show the resonance pattern for gap sizes in the surrounding 5$\times$5 patch corresponding to
 $\delta \Delta=+0.05t,0,\mbox{and} -0.05t$, respectively.
Left column (a,c,e) and right column (b,d,f) correspond to negative and positive
resonance frequency $\pm\Omega_0$, respectively.
Right panel: same as left panel but for including additional
modulations of the hopping $\delta t=-0.24t$.}\label{fig3}\end{center}
\end{figure}

From the point of view of experiment, scanning at a bias voltage
corresponding to the resonant frequency $\omega=\pm \Omega_0$ of
the Ni resonance should produce spatial patterns as shown in Fig.
\ref{fig3}.  In the left column of panels, we show the LDOS at
resonance without suppressed hopping; impurity states are hardly
observable, even in the reduced gap case.  When the hopping around
the impurity site is reduced, the pattern is clearly seen in the
suppressed gap case but not in the large gap case.  Note that the
images at positive and negative bias show fourfold patterns
which are rotated by 45$^\circ$ with respect to each other, as
observed in experiment\cite{lang} and earlier
theories.\cite{tangflatte,balatsky}

\section{ Z\lowercase{n} impurity}

The LDOS pattern predicted for an unitary scatterer in a $d$-wave
superconductor with noninteracting quasiparticles is well-known to
have crucial qualitative differences with the measured STS conductance maps
$G(eV,\rr)$, at least if these are interpreted in the usual way as
being directly proportional to the LDOS.   The primary discrepancy
is the existence of an observed intensity maximum on the central
site of the impurity pattern in the case of Zn, which is
impossible in the na\"ive theory for potentials $V_{3}$ larger
than the bandwidth, since electrons are effectively excluded from
this site. There have been several theoretical approaches to
understand this apparent paradox. The first is specific to the
STM method and relies on the fact
 that the impurity states are localized in
the CuO$_2$ plane, two layers below the BiO surface probed by the
STM tip; the intervening layers are then argued to provide a
blocking layer-specific tunneling path which samples not the Cu
directly below the tip, but preferentially the four nearest
neighbors.\cite{IMartin:2002,JXZhu:2000b} Some indirect support
for this point of view has been provided by density functional
theory,\cite{LLWang:2005} which finds that the pattern of LDOS
near the impurity but close to  the BiO surface can be quite
different than the LDOS in the CuO$_2$ plane.  On the other hand,
this calculation, applicable only to the normal state, suggested
that the hybridization of the wavefunctions involved is not only
blocking layer-specific, but also specific to the particular
chemical impurity in question.   A second class of approaches
obtains LDOS patterns similar to experiment for both Zn and Ni by
simply assuming an {\it ad hoc} distribution of site potentials
and nearby hoppings to tune the weights of on-site and
nearest-neighbor LDOS.\cite{tangflatte,JMTang:2004} While the
bare impurity potential has a much shorter range of order $\sim
1\AA$ \cite{LLWang:2005} compared to this ansatz, it is possible
that the phenomenological parameters used in these models
represent dynamically generated quantities in a more complete
theory.  Finally,  the observed Zn conductance pattern has also
been obtained in theories which describe the Zn as a Kondo
impurity\cite{vojta} and pairing impurity,\cite{andersen05}
respectively.  We now discuss  each of these scenarios in the
context of the resonance extinction problem.

\subsection{Zn as Kondo impurity}

Motivated by NMR
measurements\cite{alloulnmr,mahajan,bobroff,mendels,julien}
showing that nonmagnetic Zn impurities induce a local spin $1/2$ in their vicinity, it was proposed that  the LDOS data
could be understood within a nonlocal Kondo model.\cite{vojta,vojtabulla}
With a proper choice of a large magnetic potential coupling the
nonlocal spin associated with the impurity to the conduction
electron bath, the observed spatial pattern could be reproduced. This requires, however, the assumption of a very weak potential scattering at the Zn site.
An interpretation of this work accounted also for the
disappearance of impurity resonances in large-gap
patches\cite{vojta,hoffmanthesis} by assuming that such regions
were underdoped and therefore poorly screened.  In such a case,
the Kondo temperature $T_K$ would fall below the measurement
temperature, leaving the impurity in the local moment
(nonresonant) regime.  More recently, however, it has been
observed by STM that spatial charge variations are in fact quite
small, of order a few percent, and that the presence of dopants
{\it correlates} (rather than anticorrelates) with the large-gap
regions.\cite{mcelroyscience} In addition, resonances that have
been observed in underdoped samples at temperatures well above the
Kondo temperature expected from NMR\cite{alloul} are similar to the low temperature impurity
spectra.\cite{hudson,kambara,vershinin} Nevertheless, recently Kir\'{c}an
proposed a mechanism by which the size of the superconducting
gap can modify the impurity moment exchange coupling
to the $d$-wave quasiparticle bath.\cite{kircan} In the large-gap
regions, states are pushed further away from the Fermi level, thus
decreasing the ability of the quasiparticle system to screen the
impurity, leading to a lower $T_K$.  This does not appear to address the
set of critiques above, but is consistent with the STM results on
BSCCO at low $T$. We return to this scenario below, after
discussing other possibilities.

\subsection{Zn as screened Coulomb impurity}
\begin{figure}[t]
\includegraphics[clip=true,width=1.0\columnwidth]{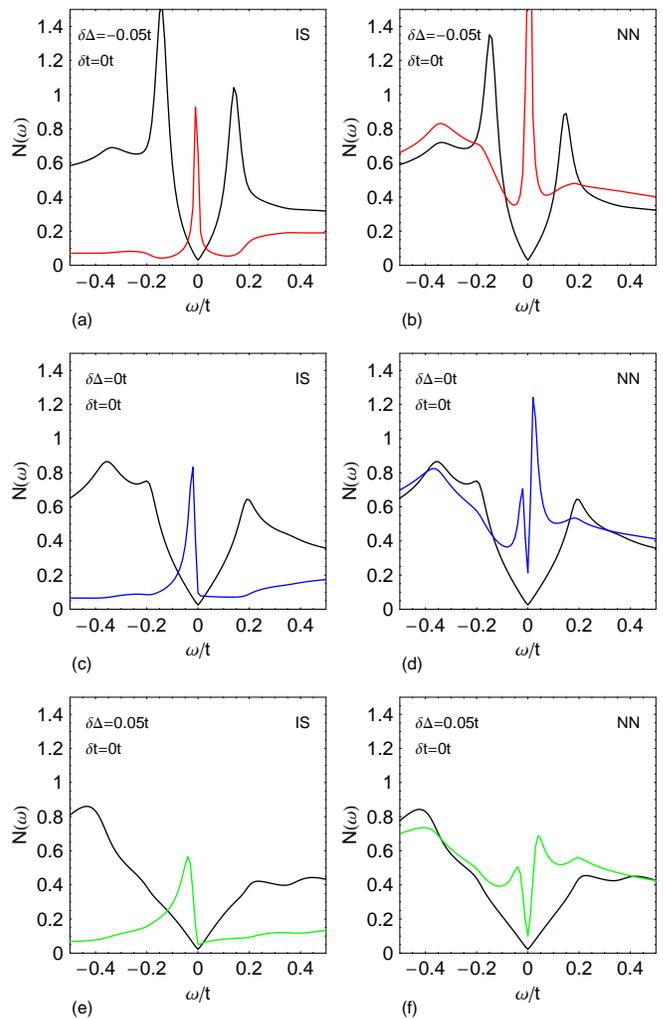}
 \caption{(Color online) LDOS at impurity site (IS) and nearest neighbor sites (NN)
 plotted in left and right panels respectively for various gap modulations
 $\delta \Delta=-0.05t,0,+0.05t$ over a 5$\times 5$ patch around the impurity, described by an
 onsite potential $V_3=2.5t$.
 Black curves show the LDOS in the absence of an impurity, while colored curves
 correspond to the situation where
 an impurity is present.  Hoppings are not modulated, $\delta t=0$} \label{fig4}
\end{figure}

We now examine  the conventional point of view that, since
Zn$^{2+}$ is a closed shell ion, it  creates a strong localized
screened Coulomb potential\cite{LLWang:2005} in a BSCCO host.  A
strong impurity potential of this type is typically represented as
a $\delta$-function potential in the Hamiltonian with $V_3>>t$ and
$V_0=0$ in the notation of Eq.(\ref{potH}), and it is well-known
that such a perturbation generates an in-gap resonant state in a
$d$-wave superconductor.\cite{bsr} In a particle-hole symmetric
normal state band, this LDOS resonance can be tuned to
$\Omega_0=0$ with $V_3=\infty$, but in a more general band a
specific fine-tuned value of the potential is required to produce
a resonance at zero energy (unitarity) or  any other particular
subgap energy.\cite{joynt,ahmphysica} For the band we have
adopted here, which roughly reproduces the correct Fermi surface
of optimally doped BSCCO, tuning the resonance to the nominal
resonance frequency observed by STS\cite{panzn} of
$\Omega_0\simeq-2$meV$\simeq 0.013t$ requires a potential of
approximately $V_3=2.5t$.  In Fig. \ref{fig4}, we have plotted the
resonance arising from such a potential on both the impurity site
(IS) and nearest neighbor site (NN), for various values of the gap
size in the local patch around the impurity, as for the Ni case.
There are two obvious difficulties. The first is the well-known
problem discussed above, that the intensity on the IS is
substantially smaller than on the NN site, in contradiction to
experiment.  The second problem relates to the behavior of the
resonance in different types of local patches.  The resonance is
suppressed somewhat in the large-gap regions corresponding to Fig.
\ref{fig4}(e-f), but is still clearly visible.  In fact, it is
clear that  the gap modulation $\delta \Delta$ has simply detuned
the resonance and acts, via its coupling through the $T$-matrix
equations to the diagonal channel, as a renormalization of the
impurity potential.  The clear splitting of the resonance and
shift of the peak with increasing $\delta \Delta$ is obvious in
panels (b,d,f).  The inability of the conventional  screened
Coulomb model for Zn to explain the real disappearance of the
resonances in the large-gap regions with this scenario was
emphasized recently by Kir\'{c}an,\cite{kircan} who studied a
similar model.

\begin{figure}[t]
\includegraphics[clip=true,width=1.0\columnwidth]{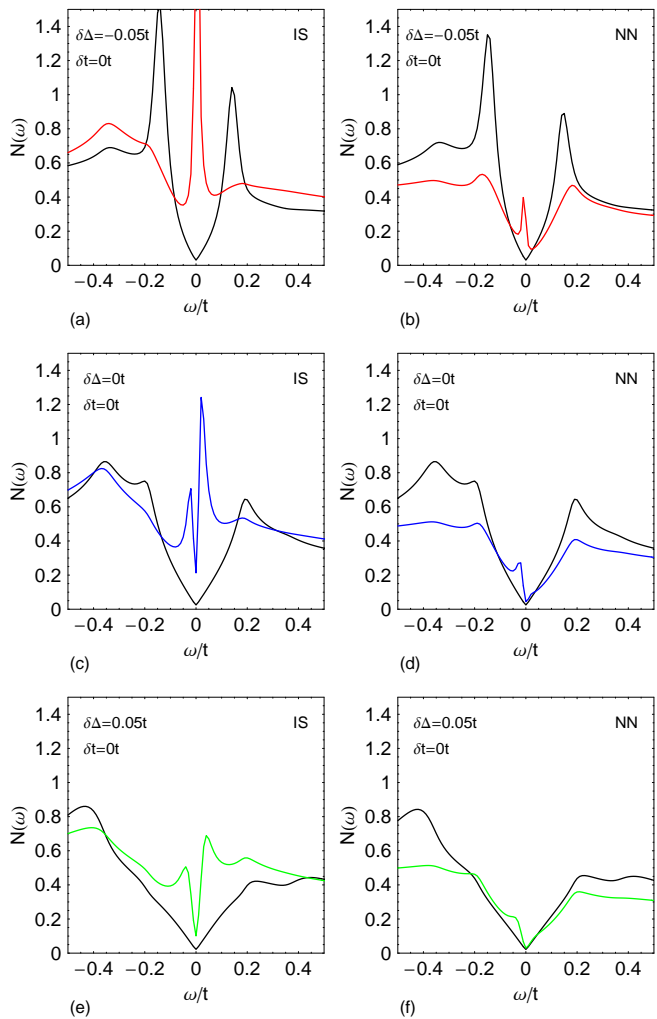}
 \caption{(Color online) Same as Fig. \ref{fig4},
but with a filter mechanism applied as described in text. } \label{fig5}
\end{figure}

We now recall that the correct spatial pattern can be recovered by
employing the assumption of a blocking layer
``filter",\cite{JXZhu:2000b,IMartin:2002} which supposes that the
tunneling path to the CuO$_2$ plane is not direct.  We employ the
simplest of these approaches,\cite{JXZhu:2000b} which assumes that
the LDOS measured on any given site is actually the average of the
LDOS on the four NN sites.  In this case, as seen from Fig.
\ref{fig5} the correct spatial pattern is
recovered,\cite{balatsky} and one sees that the resonance in the
large-gap regions is in fact hardly observable, particularly if
one searches for such objects by looking for values of the 0 meV
conductance which exceed the noise threshold, as in Ref.
\onlinecite{hoffmanthesis}. Of course, applying the same filter
mechanism to the Ni pattern will spoil the good agreement, so one
is left with the unpleasant alternative of arguing that the effect
of the intervening layers may be different in the case of the two
impurities.  This was in fact the conclusion reached by Ref.
\onlinecite{LLWang:2005}, but it destroys the utility of comparing
the patterns for the two species as a way to extract information.

\begin{figure}[t]
\begin{center}
 \includegraphics[clip=true,width=1.\columnwidth]{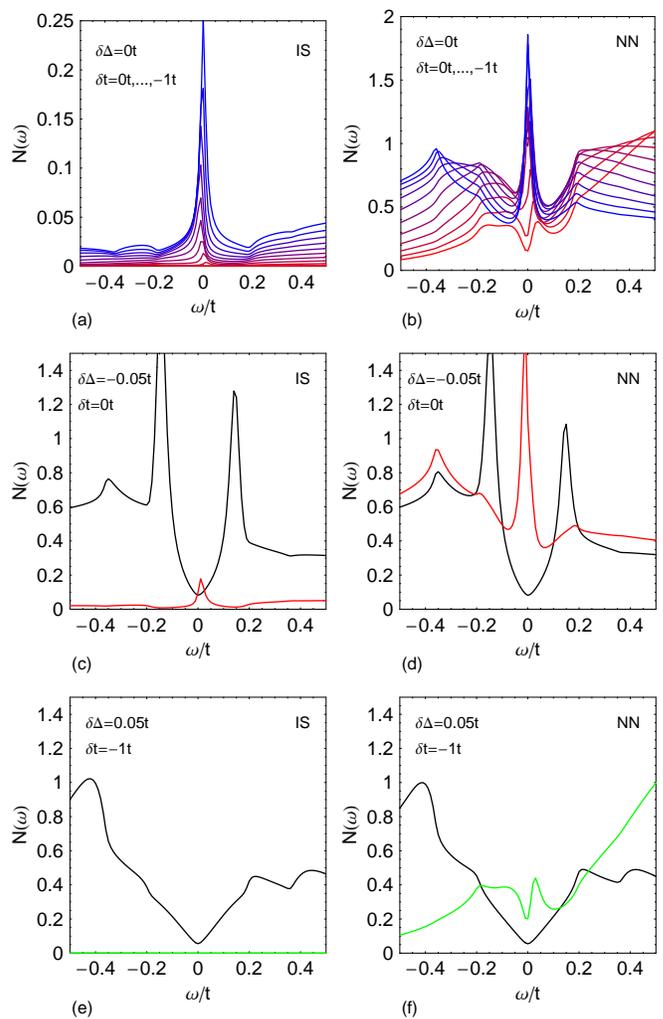}
\caption{(Color online) LDOS for a Zn impurity represented by a
potential $V_3=5t$ and calculated with a constant scattering rate
$\eta = 0.01 t$ where the hopping is suppressed using a Gaussian
distribution with the maximum $\delta t$ on the four bonds around
the impurity, and a decay length of one lattice spacing.
(a,c,e) show the LDOS at the IS site and (b,d,f) at the NN site.
(a) and (b) show the LDOS in a region of constant gap
($\delta \Delta = 0$) for different hopping suppression where the
values of $\delta t$ shown vary from 0 (top) to $-t$ (bottom) in steps of $-0.1t$.
(c) and (d) show the LDOS in a small-gap region without
suppression of the hopping while
(e) and (f) show it in large-gap region with maximum suppression
of the hopping ($\delta t = -t$).}
\label{fig6}\end{center}
\end{figure}

The resonances observed in Fig. \ref{fig5}(e) still appear as
small peaks  in the LDOS at a shifted energy, and so might be
observable; the case for such a scenario really eliminating the
resonances in the large-gap regions is therefore not completely
convincing. Before leaving this scenario, therefore, we discuss
briefly the case of reduced hopping.  We remind the reader that
due to the antiferromagnetic correlations in the host material,
the hopping is expected to be reduced over the antiferromagnetic
correlation length, of order a few lattice spacings.\cite{gabay}
This renormalization might be expected to be more important in the
case of Zn than Ni, due to the stronger bare
potential.\cite{LLWang:2005} We therefore plot in Fig.
\ref{fig6}(a,b) the resonance in a homogeneous background order
parameter ($\delta\Delta=0$) for various values of the hopping
near the impurity.  It is seen that the reduced hopping
dramatically suppresses the weight in the impurity resonance on
both the impurity [Fig. \ref{fig6}(a)] and NN [Fig. \ref{fig6}(b)]
sites, which is lost to the antibound state outside the band. The
resonance position will shift somewhat, however, since the
effective potential for the impurity which enters in the
determinant of the $T$-matrix is changing. As shown in Fig.
\ref{fig6}(c-f), assuming a sufficiently large renormalized
hopping in large-gap regions, it is possible to completely
wipe-out what would be a well-defined sharp resonance in a
small-gap region. Thus, provided the experimental data are
consistent with individual Zn's displaying small shifts in the
resonance position in intermediate strength gap patches, the
conventional scenario reviewed in the section, modified by kinetic
energy renormalization from electronic correlations, may well be
consistent with the data.

\subsection{Zn as phase impurity}

\begin{figure}[b]
\begin{center}
 \includegraphics[clip=true,width=1.\columnwidth]{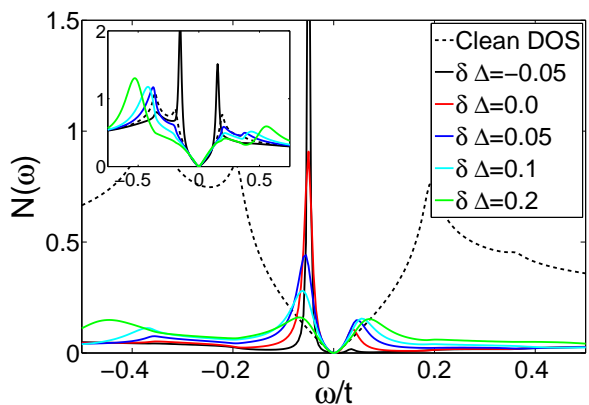}
\caption{(Color online) LDOS of a phase impurity with $\alpha=1.5$
as a function of the change in the local gap $\delta \Delta$ within the $5\times
5$ patch in which it lies. The red curve is similar to the
results in Ref. \onlinecite{andersen05}, which reported the LDOS
near a phase impurity in a homogeneous superconductor. The inset
shows the LDOS without the phase impurity (same color
code).}\label{fig7}\end{center}
\end{figure}

One of the more exotic proposals to describe the LDOS measurements
near Zn impurities is the so-called phase impurity
scenario.\cite{andersen05} As was shown in Ref.
\onlinecite{andersen05}, local sign changes of the pairing
amplitude on the four nearest bonds to the impurity site can
generate low-energy Andreev resonant states with the correct
real-space distribution (without a filter mechanism). This is true only for a sufficiently small conventional impurity potential (see Ref. \onlinecite{andersen05} for details).\cite{andersen05} Using the
same notation as in Ref. \onlinecite{andersen05}, we introduce the
parameter $\alpha$ which sets the strength of the phase impurity
by adding a potential in the $\tau_1$ Nambu channel of
$V_1=-\alpha \Delta_0$ which affects only the four NN bonds to the
impurity site in the center of the $5\times 5$ patch. As $\alpha$
increases, the Andreev resonance moves to lower energy and
sharpens up, and the resonance energy $\Omega_0$ can approach zero
when $\alpha$ becomes greater than 1.  This means that the order
parameter actually changes its phase by $\pi$ on an atomic scale.
The question relevant to the present study is how these Andreev
states depend on the local gap patch in which they are positioned.
In order to answer this, we have studied the LDOS near phase
impurities with $\alpha=1.5$ in the center of gap patches of size
$5\times 5$. Assuming that Zn causes the same local changes of the
gap irrespective of the local gap environment (i.e. same $\alpha$
for all local gap values of the $5\times 5$ patch), then one would
expect sharper resonances in small-gap regions since there the
phase impurity is effectively stronger. Indeed, the special case
of a phase impurity of strength $\alpha$ in a large-gap region
with increased local gap also of size $\alpha$ has effectively no
phase impurity at all, and hence no low-energy resonance.

In Fig. \ref{fig7} we show the LDOS at the center of the phase
impurity as a function of the local gap change $\delta \Delta$ in the $5\times 5$
gap patch. As expected, small-gap regions support well-defined
Andreev resonant states, which are however much weaker in
large-gap regions. Thus at least at a qualitative level the phase
impurity scenario seems to be fully consistent with the present
experimental STM data near Zn impurities. Note that the results in
Fig. \ref{fig7} are shown without the filter effect, and with $\delta t=0$
and a constant smearing factor $\eta=0.01t$ as opposed to an
inelastic scattering rate $\Gamma(\omega)=0.1 |\omega |$ used in
Figs. \ref{fig1}-\ref{fig5}. We find that local suppression of the hopping
mainly renormalizes $\alpha$ ($\delta t < 0$ leads to smaller effective
$\alpha$) whereas $\Gamma(\omega)$, which dominates the large-gap
regions,\cite{alldredge} obviously leads to further smearing preferentially
of the high-energy features $\omega \gtrsim \Delta$.

\section{Discussion}

Several rather general points and caveats need to be mentioned in
connection with the overall problem.  First, we have assumed that
the impurity resonances are extinguished because of some effect of
the electronic system, i.e. we ignore the possibility that the
impurities themselves are simply absent due to nonrandom processes
during the crystal growth (for a discussion, see Ref.
\onlinecite{hoffmanthesis}).

Second, we have assumed that the spectral gap is a quantity characterizing the
strength of the same electronic state in all gap patches.
Clearly, if the large-gap patches even at optimal doping correspond to a competing
electronic state, a very different approach is necessary.  The
na\"ive proposal that these gap patches correspond to pseudogap
regions does not, however, suffice to explain the absence of
impurity resonances, since most models of the pseudogap state
allow for such resonances,\cite{kruis,wang,morr,andersen032} and in addition, by heating the sample through $T_c$, they have now been
observed to survive well in to the pseudogap state.\cite{chatterjee} The Gaussian distribution of gap
sizes observed by STS also suggests that there is no qualitative
difference between, e.g.  large-gap regions and intermediate-gap
regions.

Third, we note that all our calculations (and those of others)
have been performed for models of a single impurity in some
realization of a $d$-wave superconductor.  In the real
many-impurity problem, interference between different impurity
resonances leads to distortions of fourfold symmetry and shifts of
resonance energies.\cite{stavropoulos,zah,andersen03} These
effects are particularly strong for near-unitary scatterers like
Zn.\cite{ahz03} Direct comparison of a single experimental
spectrum with a 1-impurity theoretical prediction are therefore
generally problematic.  It is likely that these effects are
responsible for the fact that the existing experimental data (for
fields of view imaged by STS containing tens of impurities) for
resonance strength versus gap size are extremely
noisy\cite{hoffmanthesis} and show no clear evidence for a
continuous suppression of the resonance peak as the local gap size
is increased, a characteristic of {\it all} the theoretical models
discussed here. Improving these statistics, with hundreds rather
than tens of impurities, and binning over a range of gap sizes,
should enable more robust conclusions to be drawn. According to
Ref. \onlinecite{ahz03}, some of the effects of interference may
also be minimized by defining the strength of the resonance via an
integral over a small energy window  near the peak rather than
simply by the value of the conductance at a fixed bias.

Fourth, both the Kondo scenario and the phase impurity scenario
produce the same spatial pattern for the Zn resonance as observed
in experiment, whereas the screened Coulomb impurity picture
requires an assumption of a filter mechanism.  If the existence or
nonexistence of such a mechanism could be settled in the case of
the small-gap regions, it would enable one to rule out at least
one candidate.  Refs. \onlinecite{andersen05} and \onlinecite{zah} 
proposed that this could be settled if one can find isolated pairs
of resonances in the same gap region.

Fifth, we discuss how the different scenarios for Zn can be
distinguished, independent of filter mechanism.   If truly
isolated resonances were available, the shift of resonance
positions from patch to patch could suffice to single out the
model with Zn as a screened Coulomb scatterer. However, as
discussed above, interference effects will likely induce such
shifts in either of the remaining scenarios as well.

The Kondo scenario for the Zn moment can be tested by a systematic
measurement of the $T$-dependence of the impurity resonances.
According to Kirc\`an,\cite{kircan} the effective Kondo scale
depends on the gap size.  Since the measured distribution of gaps
is continuous, it should be possible to find resonances in
intermediate-gap regions characterized by a small enough Kondo
scale, such that the resonance will disappear with increasing
temperature, still in the superconducting state.  In the
conventional scenarios, or in non-Kondo scenarios where the
observed moment is due to background correlations in the host
system, increasing temperature should broaden but not eliminate
the resonance until the gap disappears.

Another method to distinguish these scenarios is to apply a
magnetic field.  In a magnetic field, resonances due to an
ordinary potential will be split by a Zeeman
coupling\cite{Grimaldi} linearly in the magnetic field strength.
Splitting of the resonance peak also characterizes the phase
impurity and the Kondo resonance in a magnetic field. However, as
shown in Ref. \onlinecite{Vojtafield}, deviations from the linear
Zeeman splitting characterizes the Kondo resonance as the field
becomes of order the Kondo temperature and the two split peaks
develop a strong asymmetry in their weight. For the intermediate
gap patches with $T_K\ll T_c$ discussed above, the corresponding
field scale where the crossover occurs should be small enough such
that the unusual field dependence predicted in Ref.
\onlinecite{Vojtafield} should become observable.

Sixth,  we remind the reader that none of the models discussed
here incorporate systematically the effects of the approach
to the Mott transition on the underdoped side.  We have attempted
to include the correlation-induced band narrowing near an impurity due to 
correlations in a phenomenological way, but a more complete
treatment is clearly desirable.

Finally, we address the question of the static local moment
induced by Zn in the  underdoped regime.\cite{panagopoulos}
In the conventional model discussed above, such a moment is not
included in the theory. Recent studies however, have extended this
scenario (extended conventional scenario) by including Hubbard
correlations $U$ in the host superconductor, and found that
nonmagnetic scatterers can indeed generate--for sufficiently
strong correlations--a local $S=1/2$ state tied to the impurity
site,\cite{ting,harter} in agreement with results obtained from
the $t-J$ model treated in the slave-boson
mean-field\cite{wanglee} or Gutzwiller
approximation.\cite{tsuchiura}  An important property of the LDOS
resonance within the latter scenario is that it is intrinsically
split by the local moment whereas no such effect exists in zero
field within the Kondo approach. Therefore, experimental
observation of thermally broadened resonances which split in zero
field as the temperature is lowered would be  strong support for
the extended conventional scenario.  We stress that local
inhomogeneity may cause some resonances to exhibit this behavior
while others will not.

\section{Conclusions}

In this paper we have investigated a number of scenarios which
might explain why STS experiments fail to observe Ni and Zn
resonances in large-gap regions of the BSCCO superconductors. In
the case of Ni, it appears that the physics of an inhomogeneous
$d$-wave superconductor alone suffices to explain the phenomenon.
That is, only in the case of a coherence-length size small-gap
region embedded in a larger gap region can one expect to have
sufficiently sharp and strong features at the nominal gap
positions for the impurity resonance to borrow enough weight for
it to be well-defined.  This is a direct consequence of Andreev
states formed by partially trapped quasiparticles in suppressed
order parameter regions, as pointed out by Nunner {\it et
al.}\cite{nunner1} and Fang {\it et al.}\cite{acfang}

In discussing the similar question for the Zn resonances, we
considered several scenarios.  The first, due to Kir\'can, assumes
that the resonance is formed primarily due to Kondo screening of
the magnetic  moment formed in the correlated system around the Zn
site, and proposes that in the large-gap regions the moment is
free (unscreened).  This seems unlikely to us, given that no
temperature dependence of these resonances has  been observed
other than thermal broadening,\cite{chatterjee,kambara} but we have
proposed further tests that should clarify this issue. The second
scenario is similar to the physics of the extinction of the Ni
resonances: if the mobility of quasiparticles is reduced in the
vicinity of the Zn due to correlations, the resonance will be
suppressed as the size of the order parameter in the patch is
increased, but shifted as well. This scenario cannot presently be
ruled out by existing data. Finally, if the Zn ion primarily
influences the pair field locally, such as to cause a local $\pi$
phase shift of the order parameter, it will produce a resonance
whose weight depends directly on the size of the local gap, and
vanishes in sufficiently large gap regions.

Further theoretical and experimental  work is clearly necessary to
answer the very fundamental challenge posed here.  From the
theoretical side, we have included strong electronic correlations
only in a very phenomenological way, by introducing local
suppressions of the hole mobility around the impurity site, as
found by more sophisticated treatments in the normal
state.\cite{gabay} As these effects appear to be important,
theoretical treatments of impurities in the superconducting state
capable of treating inhomogeneous order parameter situations
together with strong correlations are clearly desirable. On the
experimental side, we have discussed ways in which improving
statistics on impurity resonances and how they are defined could
provide important insights, and proposed  various tests of the
scenarios treated here which should enable one to distinguish
them.

\section{Acknowledgements}

The authors are grateful for discussions with   J.C.~Davis, M. Vojta, and W.
Chen.  P.J.H. and S.G. were funded by DOE Grant DE-FG02-05ER46236
and S.G. acknowledges support by the Deutsche Forschungsgemeinschaft.
B.M.A. acknowledges support from the Villum Kann Rasmussen
foundation.


\begin{thebibliography}{00}
%
\bibitem{leereview} P. A. Lee, N. Nagaosa, and X.-G. Wen, Rev. Mod. Phys. {\bf 78}, 17 (2006).
%
\bibitem{sakivelson} S. A. Kivelson, I. P. Bindloss, E. Fradkin, V. Oganesyan, J. M. Tranquada, A. Kapitulnik, and C. Howald, Rev. Mod. Phys. {\bf 75}, 1201 (2003).
%
\bibitem{alvarez05} G. Alvarez,  M. Mayr, A. Moreo, and E. Dagotto, Phys. Rev. B {\bf 71}, 014514 (2005); M. Mayr, G. Alvarez, A. Moreo, and E. Dagotto, Phys. Rev. B {\bf 73}, 014509 (2006).
%
\bibitem{andersen07} B. M. Andersen, P. J. Hirschfeld, A. P. Kampf, and M. Schmid, Phys. Rev. Lett. {\bf 99}, 147002 (2007).
%
\bibitem{atkinson07} W. A. Atkinson, Phys. Rev. B {\bf 75}, 024510 (2007).
%
\bibitem{andersenkappa} B. M. Andersen and P. J. Hirschfeld, Phys. Rev. Lett. {\bf 100}, 257003 (2008).
%
\bibitem{alvarez08} G. Alvarez and E. Dagotto, arXiv:0802.3394v1.
%
\bibitem{cren} T. Cren, D. Roditchev, W. Sacks, J. Klein, J.-B. Moussy, C. Deville-Cavellin, and M. Lagu\"{e}s, Phys. Rev. Lett. {\bf 84}, 147 (2000).
%
\bibitem{howald} C. Howald, P. Fournier, and A. Kapitulnik, Phys. Rev. B {\bf 64}, 100504 (2001).
%
\bibitem{pan} S. H. Pan, J. P. O'Neal, R. L. Badzey, C. Chamon, H. Ding, J. R. Engelbrecht, Z. Wang, H. Eisaki, S. Uchida, A. K. Gupta, K.-W. Ng, E. W. Hudson, K. M. Lang, and J. C. Davis, Nature (London) {\bf 413}, 282 (2001).
%
\bibitem{lang} K. M. Lang, V. Madhavan, J. E. Hoffman, E. W. Hudson, H. Eisaki, S. Uchida, and J. C. Davis, Nature (London) {\bf 415}, 412 (2002).
%
\bibitem{panzn} S. H. Pan, E. W. Hudson, K. M. Lang, H. Eisaki, S. Uchida, and J. C. Davis, Nature (London) {\bf 403}, 746 (2000).
%
\bibitem{hudson} E. Hudson, K. M. Lang, V. Madhavan, S. H. Pan, H. Eisaki, S. Uchida, and J. C. Davis, Nature (London) {\bf 411}, 920 (2001).
%
\bibitem{tangflatte} J.-M. Tang and M. E. Flatt{\'e}, Phys. Rev. B {\bf 66}, 060504 (2002).
%
\bibitem{balatsky} A. V. Balatsky, I. Vekhter, and J.-X. Zhu, Rev. Mod. Phys. {\bf 78}, 373 (2006).
%
\bibitem{vojta}A.  Polkovnikov,  S. Sachdev, and M. Vojta, Phys. Rev. Lett. {\bf 86}, 296 (2001).
%
\bibitem{alloul} H. Alloul, J. Bobroff, M. Gabay, and P. J. Hirschfeld, arXiv:0711.0877v1.
%
\bibitem{andersen05} B. M. Andersen, A. Melikyan, T.  S. Nunner, and P. J. Hirschfeld,  Phys. Rev. Lett. {\bf 96}, 097004 (2006).
%
\bibitem{hoffmanthesis} J. Hoffman, Ph.D. thesis,
University of California-Berkeley thesis, 2003.
%
\bibitem{kruis} H. V. Kruis, I. Martin, and A. V. Balatsky, Phys. Rev. B {\bf 64}, 054501 (2001).
%
\bibitem{chatterjee} K. Chatterjee, M. C. Boyer, W. D. Wise, T. Kondo, T. Takeuchi, H. Ikuta, and E. W. Hudson,  Nature Phys. {\bf 4}, 108 (2008).
%
\bibitem{kambara}  H. Kambara, Y. Niimi, M. Ishikado, S. Uchida, and H. Fukuyama, Phys. Rev. B {\bf 76}, 052506 (2007).
%
\bibitem{davis} J. C. Davis, private communication.
%
\bibitem{kircan} M. Kir\'{c}an, Phys. Rev. B {\bf 77}, 214508 (2008).
%
\bibitem{mcelroyscience} K. McElroy, Jinho Lee, J. A. Slezak, D.-H. Lee, H. Eisaki, S. Uchida, and J. C. Davis, Science {\bf 309}, 1048 (2005).
%
\bibitem{nunner1} T. S. Nunner, B.  M. Andersen, A. Melikyan, and P. J. Hirschfeld, Phys. Rev. Lett. {\bf 95}, 177003 (2005).
%
\bibitem{nunner3} T. S. Nunner, P.  J. Hirschfeld, B. M. Andersen, A. Melikyan, and K. McElroy, Physica C, {\bf 460-462}, 446 (2007).
%
\bibitem{nunner2} T. S. Nunner,  W. Chen, B.  M. Andersen, A. Melikyan, and P. J. Hirschfeld, Phys. Rev. B {\bf 73}, 104511 (2006).
%
\bibitem{mcelroyFTSTS} K. McElroy, R. W. Simmonds, J. E. Hoffman, D.-H. Lee, J. Orenstein, H. Eisaki, S. Uchida, and J. C. Davis, Nature (London) {\bf 422}, 592 (2003).
%
\bibitem{slezak} J. A. Slezak, Jinho Lee, M. Wang, K. McElroy, K. Fujita, B. M. Andersen, P. J. Hirschfeld, H. Eisaki, S. Uchida, and J. C. Davis, Proc. Natl. Acad. Sci. USA {\bf 105}, 3203 (2008).
%
\bibitem{andersenslezak} B. M. Andersen, P. J. Hirschfeld, and J. A. Slezak, Phys. Rev. B {\bf 76},  020507 (2007).
%
\bibitem{gomes} K. K. Gomes, A. N. Pasupathy, A. Pushp, S. Ono, Y. Ando, and A. Yazdani, Nature (London) {\bf 447}, 569 (2007).
%
\bibitem{andersen06} B. M. Andersen, A. Melikyan, T.  S. Nunner, and P. J. Hirschfeld, Phys. Rev. B {\bf 74}, 060501 (2006).
%
\bibitem{loram} J. W. Loram and J. L. Tallon, arXiv:cond-mat/0609305v1.
%
\bibitem{acfang} A. C. Fang, L. Capriotti, D. J. Scalapino, S. A. Kivelson, N. Kaneko, M. Greven, and A. Kapitulnik, Phys. Rev. Lett. {\bf 96}, 017007 (2006).
%
\bibitem{stavropoulos}  D. K. Morr and N. A. Stavropoulos, Phys. Rev. B {\bf 66}, 140508 (2002).
%
\bibitem{zah}L. Zhu, W. A. Atkinson, and P. J. Hirschfeld, Phys. Rev. B {\bf 67}, 094508 (2003).
%
\bibitem{andersen03} B. M. Andersen and P. Hedeg{\aa}rd, Phys. Rev. B {\bf 67}, 172505 (2003).
%
\bibitem{gabay} M. Gabay, P. J. Hirschfeld,  E. Semel, and W. Chen,  Phys. Rev. B {\bf 77}, 165110 (2008).
%
\bibitem{alldredge} J. W. Alldredge, Jinho Lee, K. McElroy, M. Wang, K. Fujita, Y. Kohsaka, C. Taylor, H. Eisaki, S. Uchida,
P. J. Hirschfeld, and J. C. Davis, Nature Phys. {\bf 4}, 319 (2008).
%
\bibitem{JXZhu:2000b}J.-X. Zhu,  C. S. Ting, and C. R. Hu,  Phys. Rev. B {\bf 62}, 6027 (2000).
%
\bibitem{IMartin:2002} I. Martin, A. Balatsky, and J. Zaanen, Phys. Rev. Lett. {\bf 88}, 097003 (2002).
%
\bibitem{LLWang:2005}L.-L. Wang, P. J. Hirschfeld, and H.-P. Cheng, Phys. Rev. B {\bf 72},
224516 (2005).
%
\bibitem{JMTang:2004} J.-M. Tang and M. Flatt\'e, Phys. Rev. B {\bf 70}, 140510 (2004).
%
\bibitem{alloulnmr}H. Alloul, P. Mendels, H. Casalta, J. F. Marucco, and J. Arabski, Phys. Rev. Lett. {\bf 67}, 3140 (1991).
%
\bibitem{mahajan}A. V. Mahajan, H. Alloul, G. Collin, and J. F. Marucco, Phys. Rev. Lett. {\bf 72}, 3100 (1994).
%
\bibitem{bobroff}J. Bobroff, W. A. MacFarlane, H. Alloul, P. Mendels, N. Blanchard, G. Collin, and J.-F. Marucco, Phys. Rev. Lett. {\bf 83}, 4381 (1999).
%
\bibitem{mendels}P. Mendels, J. Bobroff, G. Collin, H. Alloul, M. Gabay, J. F. Marucco, N. Blanchard, and B. Grenier, Europhys. Lett. {\bf 46}, 678 (1999).
%
\bibitem{julien}M.-H. Julien, T. Feh\'er, M. Horvati\'c, C. Berthier, O. N. Bakharev, P. S\'egransan, G. Collin, and J.-F. Marucco, Phys. Rev. Lett. {\bf 84}, 3422 (2000).
%
\bibitem{vojtabulla} M. Vojta and R. Bulla, Phys. Rev. B {\bf 65}, 014511 (2001).
%
\bibitem{vershinin}M. Vershinin, S. Misra, S. Ono, Y. Abe, Y. Ando, and A.
Yazdani, Science  {\bf 303}, 1995 (2004).
%
\bibitem{bsr} A. V. Balatsky, M. I. Salkola, and A. Rosengren, Phys. Rev. B {\bf 51}, 15547 (1995).
%
\bibitem{joynt} R. Joynt, J. Low Temp. Phys. {\bf 109}, 811 (1997).
%
\bibitem{ahmphysica}W. A. Atkinson, P. J. Hirschfeld and A. H. MacDonald, Physica C {\bf 341-348},  1687 (2000).
%
\bibitem{wang} Q.-H. Wang, Phys. Rev. Lett. {\bf 88}, 057002 (2002).
%
\bibitem{morr} D. K. Morr, Phys. Rev. Lett. {\bf 89}, 106401 (2002).
%
\bibitem{andersen032} B. M. Andersen, Phys. Rev. B {\bf 68}, 094518 (2003).
%
\bibitem{ahz03}W. A. Atkinson, P. J. Hirschfeld, and L. Zhu, Phys. Rev. B {\bf 68}, 054501
(2003).
%
\bibitem{Grimaldi}C. Grimaldi, Phys. Rev. B {\bf 65}, 094502 (2002).
%
\bibitem{Vojtafield}M. Vojta, R. Zitzler, R. Bulla, and T. Pruschke, Phys. Rev. B {\bf 66}, 134527 (2002).
%
\bibitem{panagopoulos}C. Panagopoulos, J. L. Tallon, B. D. Rainford, T. Xiang, J. R. Cooper, and C. A. Scott, Phys. Rev. B {\bf 66},
064501 (2002).
%
\bibitem{ting} Y. Chen and C. S. Ting, Phys. Rev. Lett. {\bf 92}, 077203 (2004).
%
\bibitem{harter} J. W. Harter, B. M. Andersen, J. Bobroff, M. Gabay, and P. J. Hirschfeld, Phys. Rev. B {\bf 75}, 054520 (2007).
%
\bibitem{wanglee}Z. Wang and P. A. Lee, Phys. Rev. Lett. {\bf 89}, 217002 (2002).
%
\bibitem{tsuchiura}H. Tsuchiura, Y. Tanaka, M. Ogata, and S. Kashiwaya, Phys. Rev. B {\bf 64}, 140501 (2001).

\end{thebibliography}
\end{document}